\documentclass[%
 reprint,
 amsmath,amssymb,
 aps,
]{revtex4-2}

\usepackage{graphicx}
\usepackage{dcolumn}
\usepackage{bm}
\usepackage{braket}
\usepackage{mathtools}
\usepackage{tensor}
\allowdisplaybreaks

\usepackage[resetlabels,labeled]{multibib}
\newcites{App}{Appendix References}

\newcommand{\ld}{m} 
\newcommand{\nn}{\nonumber}

\usepackage{xcolor}

\begin{document}

\preprint{APS/123-QED}

\title{Generative random latent features models and statistics of natural images}

\author{Philipp Fleig}
\affiliation{Department of Cellular Biophysics, Max Planck Institute for Medical Research\\ Jahnstra{\ss}e 29, 69120 Heidelberg, Germany}
\author{Ilya Nemenman}%
\affiliation{Department of Physics, Emory University, Atlanta, GA 30322, USA}
\affiliation{Department of Biology, Emory University, Atlanta, GA 30322, USA}
\affiliation{Initiative in Theory and Modeling of Living Systems, Atlanta, GA 30322, USA}%

\date{\today}

\begin{abstract}
Complex, multivariable systems are often analyzed by grouping their constituent units into  components, sometimes referred to as latent features, which afford physical or biological interpretation. However, \textit{a priori} many different types of latent features and data decompositions can be defined, and one typically uses a trial and error approach to determine a decomposition that is natural to the system and its data. It is highly desirable to develop principled understanding of which decomposition is appropriate for given a data set. 
In this work, we take a step in this direction and argue that sample-sample correlations in the data carry important information to this effect. For this we construct a generative random latent feature matrix model of large data based on linear mixing of latent features. Key ingredient of our model is that we allow for statistical dependence between the mixing coefficients and argue that the model captures characteristic properties found in many types of natural data. Latent dimensionality and correlation patterns of the data are controlled by only two model parameters. The model's data patterns include (overlapping) clusters, sparse mixing, and constrained (non-negative) mixing. We describe the characteristic correlation and eigenvalue distributions of each pattern. Finally, we fit the model on correlation data from natural images and find a near perfect match with the sparse mixing regime of our model. This finding is in line with the well-known sparse coding structure in natural scene images and provides information about the appropriate data decomposition, namely a sparse coding scheme. We believe that our work will deliver similar insights for diverse data of biological systems.
\end{abstract}

\maketitle


\section{Introduction}

Complex physical, biological, and social systems, characterized by numerous interacting units, often exhibit emergent collective properties~\cite{anderson1972more}. To understand such systems, one typically relies on high-throughput experimental data. In such data, many variables (aka, channels or degrees of freedom, such as neural activities, mutations, or pixel values) are measured in parallel. These variables are simultaneously recorded across multiple samples, which could correspond to different time points, experimental trials, or distinct images~\cite{marre2012mapping,doi:10.1126/science.abf4588,nagy2023smart,cavagna2021novel}. Considerable success in modeling such  data has been achieved by  the detailed study of correlations between the variables of the system~\cite{schneidman2006weak,meshulam2019coarse,ruderman1997origins,halabi2009protein,Marks_etal2011,PhysRevLett.73.814}. In contrast, correlations among samples within these datasets have not been as extensively studied, and it remains less clear how they can unveil the hidden structures within the data. Nevertheless, several studies have argued that  sample-sample covariance can reveal crucial information about the data~\cite{Qin_Colwell2018,Nitzan_Brenner2021}. Similarly, statistical correlations among samples may determine the ability to learn from the sample~\cite{tomasini2024deep}.

Here, our primary focus is on analysis and modeling of  sample-sample correlations within complex datasets. To this end, we develop a generative random latent feature model to capture statistical properties of high-throughput datasets.  Remarkably, with only a handful of parameters, our model can qualitatively describe a wide variety of data patterns that can be seen in experiments. Within our model, these patterns emerge from  different ways of mixing latent features. Thus, we argue that the sample-sample correlations encode information on the nature of feature mixing and, as a result, about the suitable method for decomposing the data into features. As an example, we show how the model quantitatively describes statistical properties of image patches extracted from natural scenes, a classic dataset~\cite{olshausen_data,olshausen1996emergence}.

We consider data with $N$ variables recorded over $T$ samples. The data is large-dimensional, meaning $T, N\gg 1$. Generally, $T\sim N$, though modern experiments often push us to $N\gg T\gg1$. Surprisingly,  such large-dimensional data coming from natural systems are often simpler than they could have been in that they  reveal an intrinsically lower-dimensional, hidden structure. The number of {\it latent features} (aka, collective degrees of freedom, which may be externally imposed or emergent) present in each sample is often much smaller than $N$~\cite{Cunningham2014DimensionalityRF,gao2017theory,GALLEGO2017978,Perkins_2023,Pandarinath_etal_2018,Morrell:2021hk,nieh2021geometry,stephens2008dimensionality,halabi2009protein,moran2022defining,moran2024emergent,jordan2013behavioral}. Importantly, we note that even if the total number $\ld$ of features in the entire dataset is $\ld\sim N$ or even larger, individual data samples may still be low-dimensional, or sparse, in that only a few of the features may be combined to form  each data sample~\cite{olshausen1996emergence}.

Apart from dimensionality, the structure of data is determined by {\em how}  latent features are combined to form each datum. Here, we only explore {\em linear} mixing of features. Nonetheless, even for linear mixing, many different types of data can be observed, such as  clusters~\cite{halabi2009protein}, overlapping clusters~\cite{eisen1998cluster}, low-dimensional manifolds~\cite{goldt2020modeling}, and  sparse data~\cite{olshausen1996emergence}. The richness of low-dimensional latent feature models is further increased since some types of data have constrained values: e.~g.,  activities are often non-negative~\cite{lee1999learning}. Ideally, one generative model should be able to reproduce as many of these data structures as possible under different values of model parameters.

A realistic way to build a generative model is to rely on a random latent feature model realised as $T\times N$ data matrices with elements taken from different probability distributions. Crucially, Random Matrix Theory (RMT) can be used as a powerful tool to calculate analytical properties of such data~\cite{Potters_Bouchaud_2020}.
A number of such random latent feature models have been described in a variety of contexts~\cite{goldt2020modeling,mignacco2020role,Fleig_Nemenman_2022,Potters_Bouchaud_2020,Grosse_etal2012}. However, to the best of our knowledge, no model exists that captures the diversity of patterns between samples in the data described above. 

It is, therefore, an open question how the distribution of observable quantities like the correlations between variables and between samples, or the eigenvalues of the correlation matrix (nonzero values of which are the same for variable-variable and sample-sample correlations)\cite{Potters_Bouchaud_2020} depend on properties of the model, and in particular on its latent dimensionality and the type of feature mixing. For instance, a gap in the eigenvalue spectrum often indicates a hidden low-dimensional structure due to the low-rank nature of the data. However, as pointed out above, for the case of sparse mixing of features, overall data need not have a low-rank structure, yet it still may possess a notion of low-dimensionality. With a generative random latent feature model at hand, precisely how observable distributions depend on model properties can be studied in detail numerically, or even analytically.

Here, we present a simple random latent feature matrix model able to generate diverse types of mixing latent features. The diversity is achieved by introducing statistically dependent mixing of the features, which gives rise to distinct data patterns. We provide a qualitative discussion of the distribution of sample-sample correlations and the eigenvalue density in the different regimes of the model, and discuss the effect of feature structure on these distributions. We show that in the regime of sparse mixing of features, the model is in excellent agreement with the statistics of natural image patches and explain how this fact points to the appropriate decomposition, or coding scheme, for the data.

\section{The model}\label{sec:model}
\begin{figure*}
\includegraphics[width=17.5cm]{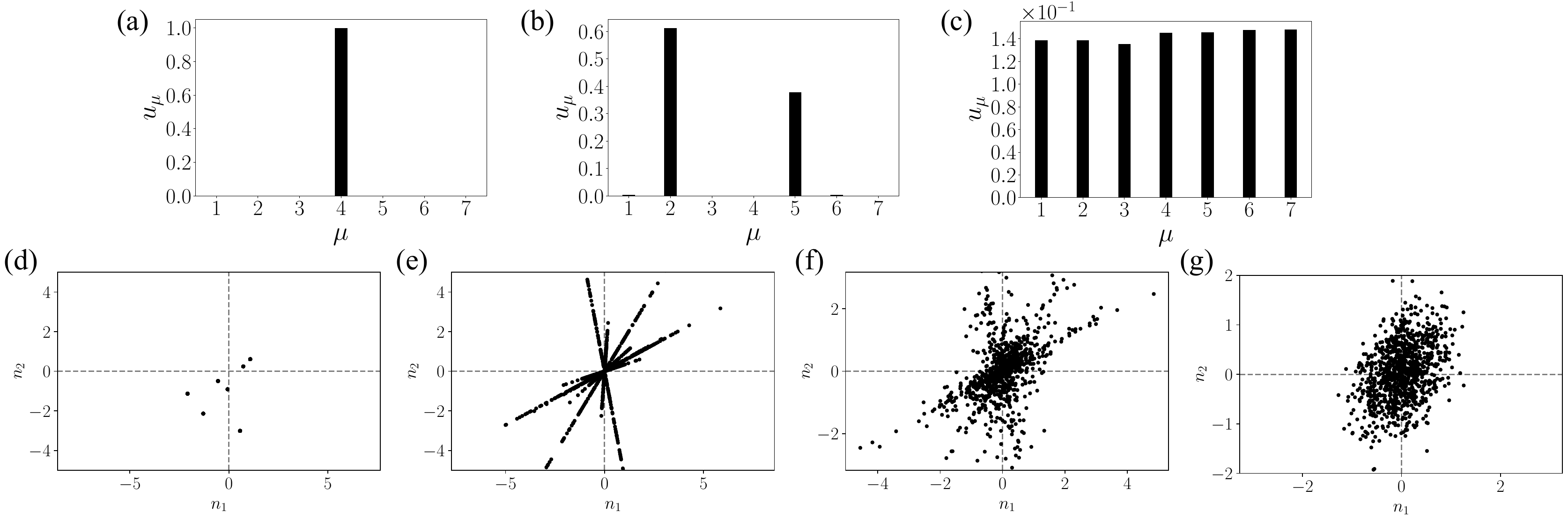}
\caption{{\bf In different parameter regimes, the SUV model admits locally low-dimensional and not low-dimensional data.} We show the data scatter of the SUV model for different choices of $\mathbf S$ and $\beta_U$. The data has a characteristic aster structure. The matrix $\mathbf V$ with random Gaussian features is the same for all three plots, and $N=2$, $T=1000$, and $\ld=7$.
(a-c) Typical samples of the Dirichlet distribution for $\beta_U=1\times10^{-8}$, $0.2$, and $1\times10^3$, respectively. (d) $s_{t\mu}=1$ (trivial modulation) and $\beta_U=1\times10^{-8}$. Each sample is given by one of the $\ld=7$ feature vectors (visible dots), creating well-defined clusters (dots), which can be further smeared by the measurement noise. (e) $s_{t\mu}\sim \mathcal N(0,1)$ and $\beta_U=1\times10^{-8}$. Each sample is still associated with a single feature vector, but the activation strength of the features follow the Gaussian distribution. This creates data that are locally low-dimensional, even though the overall dimensionality of the data cloud is not low. (f) $s_{t\mu}\sim \mathcal N(0,1)$ and $\beta_U=0.2$. Each sample is given by a mixture of multiple feature vectors (on average $m\beta_U=1.4$ features) and the locally low-dimensional structure, while still visible, is not as prominent. (g) $s_{t\mu}\sim \mathcal N(0,1)$ and $\beta_U=1000$. On average, each sample is given by a mixture of all feature vectors and the local low-dimensional structure has disappeared.}\label{fig:fig1}
\end{figure*}
We consider data matrices $\mathbf X$ with $N$ variables (columns) recorded in $T$ samples, or observations (rows). We focus on models where $\mathbf X$ is the product of three random matrices of the following form (note that $\circ$ denotes element-wise multiplication):
\begin{align}\label{eq:model}
    \mathbf X &= (\mathbf S \circ\mathbf U)\mathbf V \mbox{ or, for the matrix elements,}\\
        x_{tn}&=\sum_{\mu=1}^\ld s_{t\mu}u_{t\mu} v_{\mu n}\,.
\end{align} 
We refer to this as the \textit{SUV} model and denote matrix elements by lower case letters throughout this article.
We call $\mathbf V$ the {\em latent feature matrix} (dimensions $\ld\times N$). Each row of the matrix is the value of one of $\ld$ latent features, each specified as a row vector across the $N$ variables. Further, $\mathbf U$ is the {\em coefficient matrix} (dimensions $T\times \ld$), which measure how ``active'' each of the latent features is in each sample. Finally,  $\mathbf S$ (dimensions $T\times\ld$) is the {\em modulation matrix}, which can change the contribution of a feature to a sample by amplifying, zeroing it out, changing its sign, etc. Thus, overall, the product $\mathbf S\circ\mathbf U$, the {\em mixing matrix}, describes how each of the $T$ samples is obtained as a mixture of $\ld$ latent features. The reason for representing the mixing matrix as a product of two factors will become clear later. Briefly,  this is because products of two factors, each with simple statistical properties,  can produce data  with complex, natural features. 

As examples of data, which we hope to represent by this model, consider first image analysis. Here, one may considers $T$ samples as images in a dataset, each image consists of $N$ pixels. The  latent features are $\ld$ basis images, which are linearly combined to construct an image in the sample. Recordings of neural population activity are another example of data that we would like to represent. Here the variables are the  $N$ recorded neurons, with $T$  samples of activity of this population. The latent features represent $\ld$ modes of collective neural activity \cite{GALLEGO2017978,Pandarinath_etal_2018,Perkins_2023}. We will assume that $T,N\gg 1$, which is often the case for such data.

In~\cite{Fleig_Nemenman_2022}, we considered possibly the simplest  model of the SUV structure, namely $\mathbf X=\mathbf U\mathbf V$ (with additional additive noise) for the case when the elements of $\mathbf U$ and $\mathbf V$ were i.~i.~d. Gaussian random variables, and the elements of $\mathbf S$ where all ones. We summarise the details of the model in Appendix~\ref{app:Gaussian-Gaussian}. In this model, each sample is a linear combination of $\ld$ features, and the coefficients are independent Gaussian random variables themselves. Thus sample correlations are introduced by statistically independent mixing of the $\ld$ latent features. Going forward, we will  refer to this model as the \textit{Gaussian-Gaussian} (GG) model. Here, we consider more complicated SUV models by dropping the condition of statistical independence of the mixing matrix components. Specifically, we are seeking to introduce mixing matrices that can model sparsity, often seen in biological data \cite{OLSHAUSEN19973311}, so that only a few of many features can contribute to data at any given time. We propose to model this by viewing the elements of $\mathbf U$ as taken from a Dirichlet distribution \cite{NIPS2001_d46e1fcf}:
\begin{align}\label{eq:Dirichletdistr}
    {\rm Dir}(\{u_\mu\}; \beta_U)=\frac{1}{B(\beta_U)}\prod_{\mu=1}^\ld u_\mu^{\beta_U-1}\,,\quad \sum_\mu u_\mu=1\,,
\end{align}
where $\beta_U\in\mathbb R_{>0}$ is a hyperparameter,  $B(\beta_U)$ is 
\begin{align}
    B(\beta_U)= \frac{ \Gamma^m(\beta_U)}{\Gamma(m\beta_U)}\,,
\end{align}
and $\Gamma(\beta_U)$ is the usual Gamma-function of its argument. The Dirichlet distribution generates random vectors of $\ld$ weights $\{u_\mu :\mu=1,\ldots,\ld\}$, with the weights summing to unity. If the rows of the coefficient matrix are Dirichlet distributed, then (positive) contributions of different latent features to the observations are all dependent. While, in principle we can consider complex feature matrices $\mathbf V$ (which may be important, for instance, for generating data that is required to be strictly non-negative), for most of this article, we continue to model the features themselves as Gaussian distributed:
\begin{equation}\label{eq:DG_model}
    v_{\mu n}\sim \mathcal N(\mu_V,\sigma_V^2)\,.
\end{equation}

To develop some intuition for what follows, we note that the mean, the variance, and the covariance of weights in the random vector sampled from the Dirichlet distribution are easy to calculate~\cite{DirichletWiki} \begin{align}
    \mu_u\equiv\bar{u}_\mu&=\frac{1}{\ld}\,,\\
     \sigma_u^2\equiv\text{var}\,{u_\mu}&=\frac{1-\ld^{-1}}{\ld(1+\ld \beta)}\,,\label{eq:Dirichlet_mean&variance}\\
    \text{cov}(u_\mu,u_\nu)&=\frac{\delta_{\mu\nu}-\ld^{-1}}{\ld(1+\ld\beta)}\label{eq:Dirichlet_covariance}\,.
\end{align}

Entries of $\mathbf U$ are limited to $[0,1]$, which is too restrictive to allow agreement with real data. Thus, we use $\mathbf S$, which modulates the coefficient matrix by multiplying each of its entries by a random number.  We take the elements of $\mathbf S$ to be statistically independent, so that all correlations between the entries of the mixing matrix are in the coefficients, and not in their modulation. In Fig.~\ref{fig:fig1}, we illustrate the typical data scatter of the SUV model in three examples with $\ld=7$. In Fig.~\ref{fig:fig1}(d), we set all elements $s_{t\mu}=1$, such that the modulation of $\mathbf U$ is trivial. For small $\beta_U$ (e.g., $\beta=1\times 10^{-8}$), each sample is given by a single feature vector, and we can thus expect to see exactly $\ld$ points in the scatter plot. In contrast, in Fig.~\ref{fig:fig1}(e), the components of $\mathbf S$ are i.~i.~d. random variables drawn from a Gaussian distribution. For small $\beta_U$, still only a single feature is associated with each sample, but the activation strength of features varies, producing a characteristic aster scatter of samples. These data are locally low-dimensional, but globally they are not. That is, in each single petal of the aster in Fig.~\ref{fig:fig1} (e), the data is locally one-dimensional, while the whole aster has the dimensionality of two. In contrast, in Fig.~\ref{fig:fig1} (f), $\beta_U=0.2$, and multiple features contribute at different activation strengths, so that the one-dimensional structure of the aster has been broken (though remnants of it are still visible), and each petal  now has the same dimensionality as the whole aster. Finally, in Fig.~\ref{fig:fig1} (g), $\beta_U=1000$, so that on average, all features contribute to a sample and no locally low-dimensional structure is left. In our application to natural image data in Sec.~\ref{sec:natural_images}, we explore yet another choice of $\mathbf S$, where the matrix elements are binary i.~i.~d. random variables.

Data analysis usually focuses on correlations among the measured variables.  Since different such variables may have widely different units, one often uses empirically standardised data $\mathbf X'$, where each variable (columns of the data matrix $\mathbf X$) is independently centered to zero mean and scaled to unit variance. One then makes inferences about the data based on the empirical variable-variable  (denoted by subscript $\mathrm{v}$) correlation matrix, 
\begin{align}
    \mathbf C_\mathrm{v}=\frac 1T\mathbf X'^T\mathbf X'\,.
\end{align}
In contrast, in this work, we focus on the correlations between samples. We also work with empirically standardised data $\widetilde{\mathbf X}$, where now each sample (row of the data matrix $\mathbf X$) is independently centered to zero mean and scaled to unit variance.  The empirical sample-sample correlation matrix (denoted by subscript $\mathrm{s}$) is 
\begin{align}\label{eq:corr_matrix}
   \mathbf C_\mathrm{s}=\frac1N\widetilde{\mathbf X}\widetilde{\mathbf X}^T\,.
\end{align}
The empirically standardised data approximate the {\em theoretically} standardised random variables
\begin{align}\label{eq:x_standardised}
    \widetilde x \equiv \frac{x-\mu_x}{\sigma_x}\,,
\end{align}
where the mean $\mu_x$ and the standard deviation $\sigma_x$ of the pdf of each element $x$ of $\mathbf X$ (also denoted as {\em theoretical} values) are obtained from Eq.~(\ref{eq:model}):
\begin{align}
    \mu_x=\mathbb E[x]&=\sum_\mu\mathbb E[s_\mu u_\mu v_\mu]=\sum_\mu\mathbb E[s_\mu]\mathbb E[u_\mu]\mathbb E[v_\mu]\,,\label{eq:mean_x}\\
    \sigma^2_x&=\mathbb E[(x-\mu_x)^2]=\sum_\mu\text{var}(s_\mu u_\mu v_\mu)
    \nonumber\\&+\sum_{\mu\neq\nu}\text{cov}\left(s_\mu u_\mu v_\mu, s_\nu u_\nu v_\nu\right)\,.\label{eq:var_x}
\end{align}
Above we used the fact that $s_\mu$, $u_\mu$, and $v_\mu$ are statistically independent from each other. 

We will study the eigenvalues of the sample-sample correlation matrix $\mathbf C_\mathrm{s}$, denoted  $\lambda_\mathrm{s}$. However, we note that there is a well-known relation  between the eigenvalues of $\mathbf C_\mathrm{s}$ and  eigenvalues 
of $\mathbf C_\mathrm{v}$, denoted  $\lambda_\mathrm{v}$: nonzero eigenvalues of both matrices differ only by a multiplicative constant~\cite{Potters_Bouchaud_2020}. Specifically, it is easy to see that  
\begin{align}
    \lambda_\mathrm{v}\approx\frac{\bar\sigma_\mathrm{s}^2N}{\bar\sigma_\mathrm{v}^2T}\lambda_\mathrm{s}\,,
\end{align}
where $\bar\sigma_\mathrm{s}$ is the empirical standard deviation along the dimension of variables (the bar denotes that we take the average value of all samples), and $\bar\sigma_\mathrm{v}$ is the averaged empirical standard deviation along the dimension of samples (averaged over variables) of $\mathbf X$. For $T,N\gg 1$, $\bar\sigma_\mathrm{s}$ and $\bar\sigma_\mathrm{v}$ both converge to the theoretical value $\sigma_x$ and $\bar\sigma_\mathrm{s}/\bar\sigma_\mathrm{v}\rightarrow1$.
In what follows, we focus on the $N,T\gg1$ regime, so that the row and column sample variances are approximated by their theoretical  values. 

\section{Data patterns and  distributions of observables}
\begin{figure*}
\includegraphics[width=16.4cm]{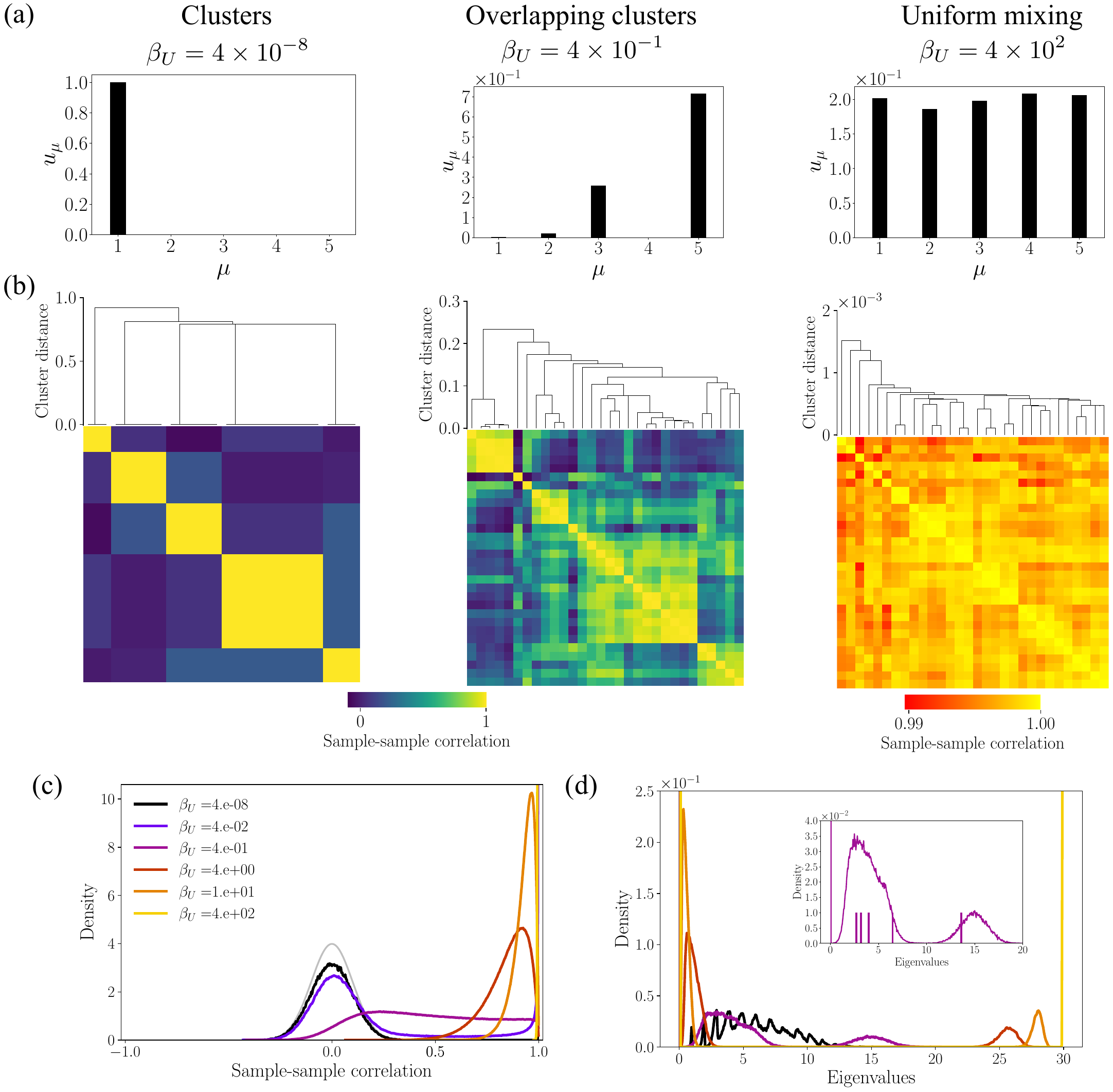}
\caption{{\bf Qualitatively different correlation patterns emerge in the Dirichlet-Gaussian model in different regimes of the parameter $\beta_U$:} We illustrate this for $T=30$, $N=100$, $\ld=5$, and $\mathbf V$ has i.~i.~d. Gaussian entries with $\sigma_V^2=1$. (a) Typical samples of the Dirichlet distribution for different $\beta_U$. (b) Sample-sample correlation matrices and dendrograms obtained from hierarchical clustering of the ECM. (c) Probability densities of correlation coefficients for a range of $\beta_U$ values. For reference, the gray curve shows the Gaussian distribution with the variance $1/N$, as expected for correlations of pure Gaussian noise data. (d) Eigenvalue densities for the same $\beta_U$ values. Inset: zoom in for $\beta_U=4\times 10^{-1}$, which is in the overlapping clusters regime. The vertical bars show the positions of eigenvalues for the single realization of the model; the height of bars is not meaningful. Note a single large eigenvalue separated from the rest by a gap. Averages are computed from $2\times 10^4$ independent realizations of the model.}\label{fig:fig2}
\end{figure*}
To explore the effects that statistically dependent mixing of features has on the structure of data, we first choose a trivial modulation of the matrix $\mathbf U$ by setting all elements of $\mathbf S$ to unity:
\begin{align}
    s_{t\mu}=1\text{ for all $t,\mu$}\,.
\end{align}
Furthermore, we again take the entries of the feature matrix $\mathbf V$ as i.~i.~d. Gaussian random variables. We will refer to this model as the \textit{Dirichlet-Gaussian} model. For convenience, we will set $\mu_V=0$ and $\sigma_V^2=1$ in the following, unless explicitly stated otherwise.

\subsection{Data patterns}
\begin{figure*}
\includegraphics[width=17.4cm]{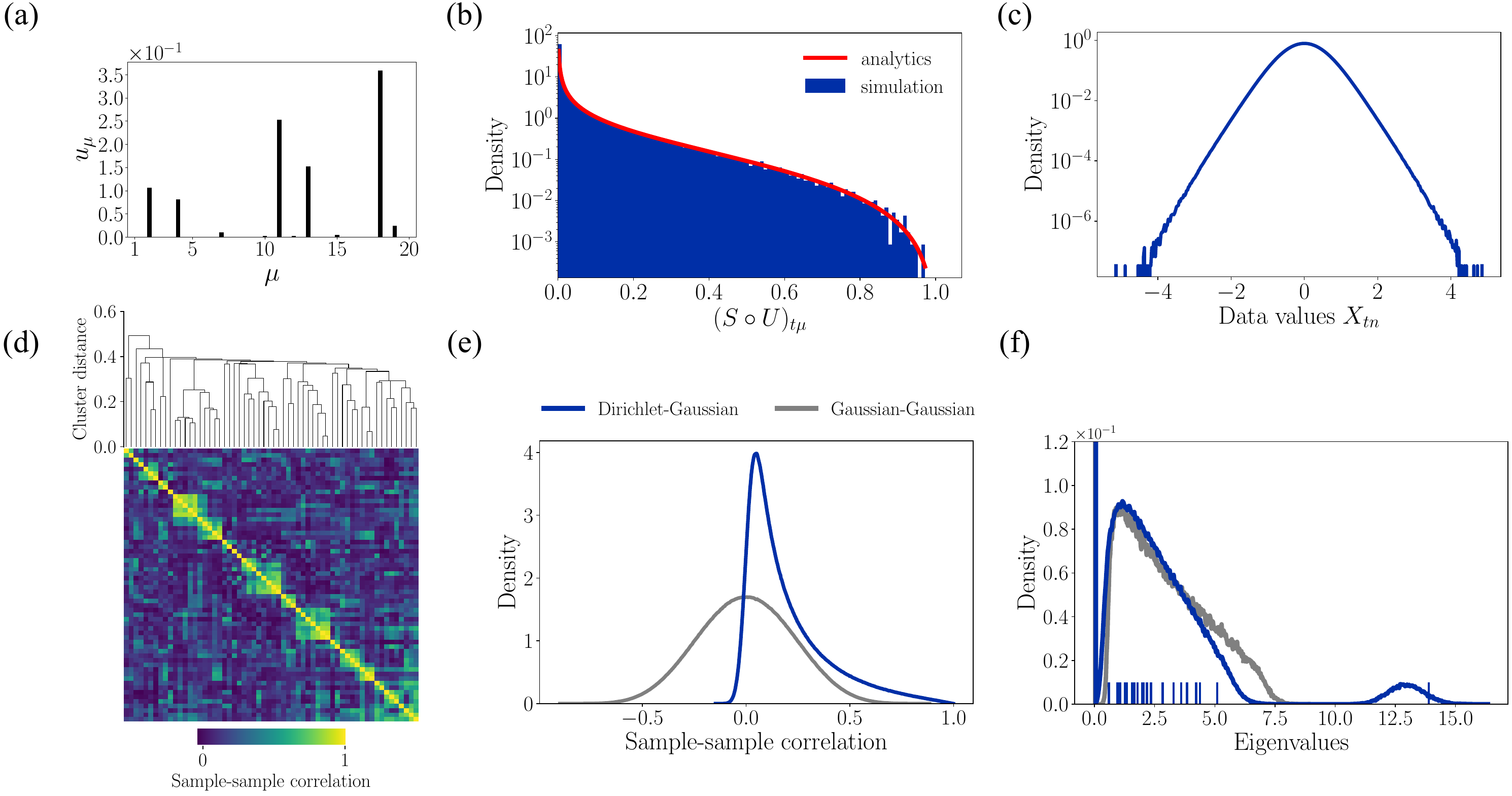}
\caption{{\bf The SUV model admits a sparse mixing regime.} We use $T=60$, $N=1000$, $\ld=20$, and $\sigma_V^2=1$. (a) Typical samples of the Dirichlet distribution for $\beta_U=0.15$. (b) Distribution of entries of the mixing matrix. Simulation from $100$ independent samples (blue), and analytics (red). (c) Distribution of data values. (d) Sample-sample correlation matrix and dendrogram obtained from hierarchical clustering of the ECM. (e) Density of sample-sample correlations of the Dirichlet-Gaussian model (blue). The observable density of the Gaussian-Gaussian model (gray) is shown for reference.
(f) Eigenvalue densities of the same models. Vertical bars show  positions of eigenvalues for the single realization of the Dirichlet-Gaussian model; the height of bars is not meaningful. Note a single large eigenvalue separated from the rest by a gap. Averages are computed from $2\times 10^4$ independent model realizations.}\label{fig:fig3}
\end{figure*}

We give a description of the data patterns obtained in different regimes of the parameter $\beta_U$. We consider both the case when $\ld\sim O(1)$ and when $\ld\sim O(N)$. In Figs.~\ref{fig:fig2} (a) and~\ref{fig:fig3} (a), we show the  typical draws from the Dirichlet distribution, and in Figs.~\ref{fig:fig2} (b) and~\ref{fig:fig3} (d), we visualise data patterns by presenting the pairwise correlation of the samples for these draws. We plot the correlation matrices with  samples ordered  using the nearest neighbour clustering~\cite{SciPy_nearest_neighbour}, which group the samples hierarchically based on their correlations. The dendrograms above the correlation matrices illustrate the hierarchical clustering. Below we describe typical patterns observed in the data.

\emph{Clusters}, Fig.~\ref{fig:fig2} (left column), emerge for $\ld\sim O(1)$ and $\beta_U\ll 1$. Here each column of the coefficient matrix $\mathbf U$ is concentrated on a single feature, and each feature defines a separate cluster. The clusters are clearly visible in the correlation matrix. The dendrogram shows that the cluster are clearly separated, and each cluster includes multiple samples.

\emph{Overlapping clusters}, Fig.~\ref{fig:fig2} (middle), are visible for $\ld\sim O(1)$ and $\ld\beta_U\sim O (1)$. Here, $O(1)$ features contribute to each sample. The correlation structure between samples becomes fuzzy, and samples may belong to more than one cluster. The dendrogram reveals groups of samples of different sizes and varying cluster distances.

\emph{Uniform mixing}, Fig.~\ref{fig:fig2} (right), describes the regime $\ld\sim O(1)$ and $\beta_U\gg 1$.  All of the $\ld$ latent features contribute approximately uniformly to each sample, and all the samples are, therefore, highly correlated. The dendrogram confirms that there is effectively a single cluster of samples (note that the scale of cluster correlation distances is $10^{-3}$ in this subplot).

\emph{Sparse mixing}, Fig.~\ref{fig:fig3}, is observed for $\ld\gg 1$, and $\ld\beta_U\ll \ld$ ($\ld\beta_U$ can be $O(1)$, or $\gg1$). In this  regime, the number of features available is much larger than the number of features that contribute to any given sample.  In Fig.~\ref{fig:fig3}(b), we show the distribution of elements of the simulated mixing matrix (blue) and an analytic curve (red) given by Eq.~(\ref{eq:mixing_matrix_distribution}). The plot shows a large number of small entry values and a small number of large entries, as expected for sparse mixing. For example, in  Fig.~\ref{fig:fig3}, we chose $\ld=N/3$ and, on average, there are $\ld\beta_U=3$ features per sample. In Fig.~\ref{fig:fig3}(c) we show the distribution of data matrix entries. From the dendrogram, Fig.~\ref{fig:fig3}(d), we see that there are no clear clusters of more than two samples.

\subsection{Observable distributions}

Next, we qualitatively discuss the distribution of correlation coefficients and the spectral density of the ECM, Eq.~(\ref{eq:corr_matrix}). In the simulations,  the pairwise correlations between variables are computed from empirically standardized data. These approximate the theoretically standardized random variables in Eq.~(\ref{eq:x_standardised}), whose means and variances we now compute.

The mean of a product of a zero mean Gaussian and a Dirichlet random variable vanishes. Hence, by Eq.~(\ref{eq:mean_x}), the mean of elements $x$ of the data matrix also vanishes
\begin{align}
    \mu_x=\sum_\mu \mathbb E[v_\mu]\mathbb E[u_\mu]=0\,.
\end{align}
In Appendix~\ref{app:variance_GaussDirichlet} we evaluate Eq.~(\ref{eq:var_x}) to find the variance:
\begin{align}
    \sigma^2_x=\ld(\sigma_u^2+\ld^{-2})\sigma_V^2=\left(\frac{1-\ld^{-1}}{1+m\beta_U}+\frac{1}{m}\right)\sigma_V^2\,.
\end{align}
We verify this expression numerically in Fig.~\ref{fig:S1}(a), and note the following limits: when $\ld\gg 1$, while $\ld\beta_U$ is finite, the variance approaches $\sigma^2_x\approx \sigma_V^2/(1+\ld\beta_U)$; when $\beta_U$ approaches zero, $\sigma^2_x\approx \sigma_V^2$; and when $\beta_U$ approaches infinity, $\sigma^2_x\approx \sigma_V^2/\ld$.

\subsubsection{Correlations}
The family of correlation distributions in Fig.~\ref{fig:fig2}(c) reflects the transition between the two extreme limits (from pure clusters to uniform mixing). The first observation is that the density is not symmetric around zero. In the clusters limit (black curve), the density of correlations has a nearly delta function peak at 1, corresponding to correlations between samples which are contributed to by the same features. The density also has  an approximately Gaussian peak around zero, which emerges from correlations between samples that include different features, and hence from feature-feature correlations. In the opposite limit of uniform mixing (yellow curve), all small correlations have disappeared and the density is approaching a delta function concentrated at 1 because all variables approximately behave as a single cluster. For the regime of overlapping clusters between the two extremes, we observe  intermediate correlation values between zero and one, signifying the existence of variables that can be attributed to more than one cluster. 

In the sparse mixing regime, the correlation density (blue curve in Fig.~\ref{fig:fig3}(e)) is again non-symmetric around zero. It has a large peak around small, positive correlations and few large correlation values. This is due to the sparse selection from a large number of features, so that there are now samples with different numbers of the same features, smearing the high correlation peak of the clusters limit. The peak near zero still comes from correlations between samples that do not have the same features, and hence from feature-feature correlations. For reference, we also show the correlation density of a Gaussian-Gaussian random latent feature model (gray curve). In the Gaussian-Gaussian model, the feature and the coefficient matrix have i.~i.~d. zero mean, unit variance Gaussian entries. The density of the Gaussian-Gaussian model was analyzed in detail in~\cite{Fleig_Nemenman_2022} and for the reader's benefit we provide a very brief summary of the model in Appendix~\ref{app:Gaussian-Gaussian}.

\subsubsection{Eigenvalues}
The shape of the eigenvalue density depends on the model parameters $T$, $N$, $\ld$ and $\beta_U$. From RMT, we know that the eigenvalue densities of random matrices often converge to their limiting form when $T$, $N$, etc.\ tend to infinity, while ratios, such as $N/T$ or the ratio of the number of latent features to the number of observables $\ld/N$ remain fixed. This is known as the thermodynamic or large-$N$ limit, in which the fixed  ratios control the shape of the density~\cite{Potters_Bouchaud_2020,Fleig_Nemenman_2022}.  We  expect these ratios to control the shape of the eigenvalue distributions of the Dirichlet-Gaussian model as well. However, here we focus on the qualitative description of eigenvalue densities, and we leave a detailed analytic description for the future.

When $\ld\sim O(1)$, the rank of the ECM satisfies $\mathrm{rank}(\mathbf C)<N,T$, implying that there are trivial zero eigenvalues. Beyond these, analogous to the correlations, the eigenvalue densities in Fig.~\ref{fig:fig2}(d) reflect the transition between the two limits of the model.

First, in the clusters limit (black curve), the eigenvalue density has a broad, `spiky' bump. The spikes result from the interaction of (on average) $\ld$ dominant eigenvalues of a similar magnitude. In Fig.~\ref{fig:S2}, we show how the resolution of the delta spikes increases at better sampling (smaller $T/N$ ratio). In the opposite limit of uniform mixing (yellow curve), the spectrum becomes degenerate with two delta peaks. One peak is for a single eigenvalue at $\lambda=N$, corresponding to all variables being in one cluster, and the other peak has $N-1$ eigenvalues at zero.

For the regime of overlapping clusters, in between the two limits, the density has two bumps (inset in Fig.~\ref{fig:fig2} (d)). The bumps move in opposite directions as $\beta_U$  increases, and they develop into delta peaks in the extreme limit. To understand the origin of the two bumps, we show the position of eigenvalues of a particular realization of the overlapping clusters model from Fig.~\ref{fig:fig2} (b) by vertical bars in Fig.~\ref{fig:fig2} (d, inset). The right density bump is due to a single, top-ranked eigenvalue, and the left bump is due to the remaining $\ld-1$ non-trivial eigenvalues. In the particular case shown, the left bump is the distribution of four eigenvalues. We verified numerically our expectation that the left bump carries approximately four times as much mass as the right bump. The top-ranked eigenvalue is due to a baseline correlation across all variables that comes from the non-zero, positive mean of the correlation density. Its origin can be further understood from analysing the eigenmode associated with the top-ranked eigenvalue. For this, we decompose the correlation matrix into two contributions, as expressed in Fig.~\ref{fig:S3} (a). The first contribution is the projection of the correlation matrix onto the top-ranked eigenmode, and the second contribution is the projection on all other eigenmodes. The decomposition is shown in Fig.~\ref{fig:S3} (b), and the distributions of matrix entries of each contribution is shown on the right. The mean values of the correlation density and the top-ranked contribution are the same, and the mean of the remaining contribution vanishes. Thus the top-ranked mode can be considered as a baseline contribution, similar to the `market mode' found in the analysis of financial data, which captures the collective up-and-down of the stock market~\cite{Laloux_etal_1999}. As $\beta_U$ grows, more and more of the total variance goes into this mode, until all variables are strongly correlated. The remaining $\ld-1$ modes capture correlations between specific subgroups of variables. As $\beta_U$ is increased, less and less of the total variance belongs to these modes.

Finally, in Fig.~\ref{fig:fig3}(f) we show the eigenvalue density in the {\em sparse} mixing regime (blue curve) and for reference we also show the eigenvalue density  of the Gaussian-Gaussian model (gray curve) \cite{Fleig_Nemenman_2022}. The density of the Gaussian-Gaussian model has a single peak with upper and lower eigenvalue bounds given by $\lambda_\pm=(1\pm \sqrt{N/\ld})^2\approx0.54,\,7.46$, respectively, as was derived in \cite{Fleig_Nemenman_2022}. In contrast, the eigenvalue density of the sparse mixing model has two peaks. The qualitative understanding of the origin of the two peaks is analogous to the one just described for the overlapping clusters regime. The eigenvalues of a specific realization of the sparse Dirichlet-Gaussian model are shown by the blue bars here. The right and the left bump, are due to a single top-ranked and the $\ld-1$ remaining non-trivial eigenvalues, respectively. The decomposition of the sparse Dirichlet-Gaussian correlation matrix into the two contributions is shown in Fig.~\ref{fig:S3} (c), together with the contributions' densities.

For the computation of the sparse model we have selected parameters such that the density is well sampled. This is the case when both $T$ and $N$ are large. In particular, $N$ has to be large enough to ensure that sampling of the different ways of sparsely selecting features is sufficient. In other sampling limits, the form of the eigenvalue density is more complicated, and a detailed analysis will be needed to understand its structure.

Finally, in Fig.~\ref{fig:S4} we show the influence that feature structure has on the correlation and eigenvalue observable distributions for the sparse mixing regime. As examples, we consider features which are Dirichlet random vectors, features with entries that are i.~i.~d. random variables drawn from an exponential distribution, and features with entries that are i.~i.~d. binary random variables taking values $\{\pm1\}$. We note that the first two examples lead to non-negative data  and the SUV model is a form of non-negative matrix factorization~\cite{lee1999learning} in these cases. We observe that feature structure only has a small influence on the correlation and eigenvalue distributions.

\section{Sample statistics of natural images}\label{sec:natural_images}
\begin{figure*}
\includegraphics[width=17.4cm]{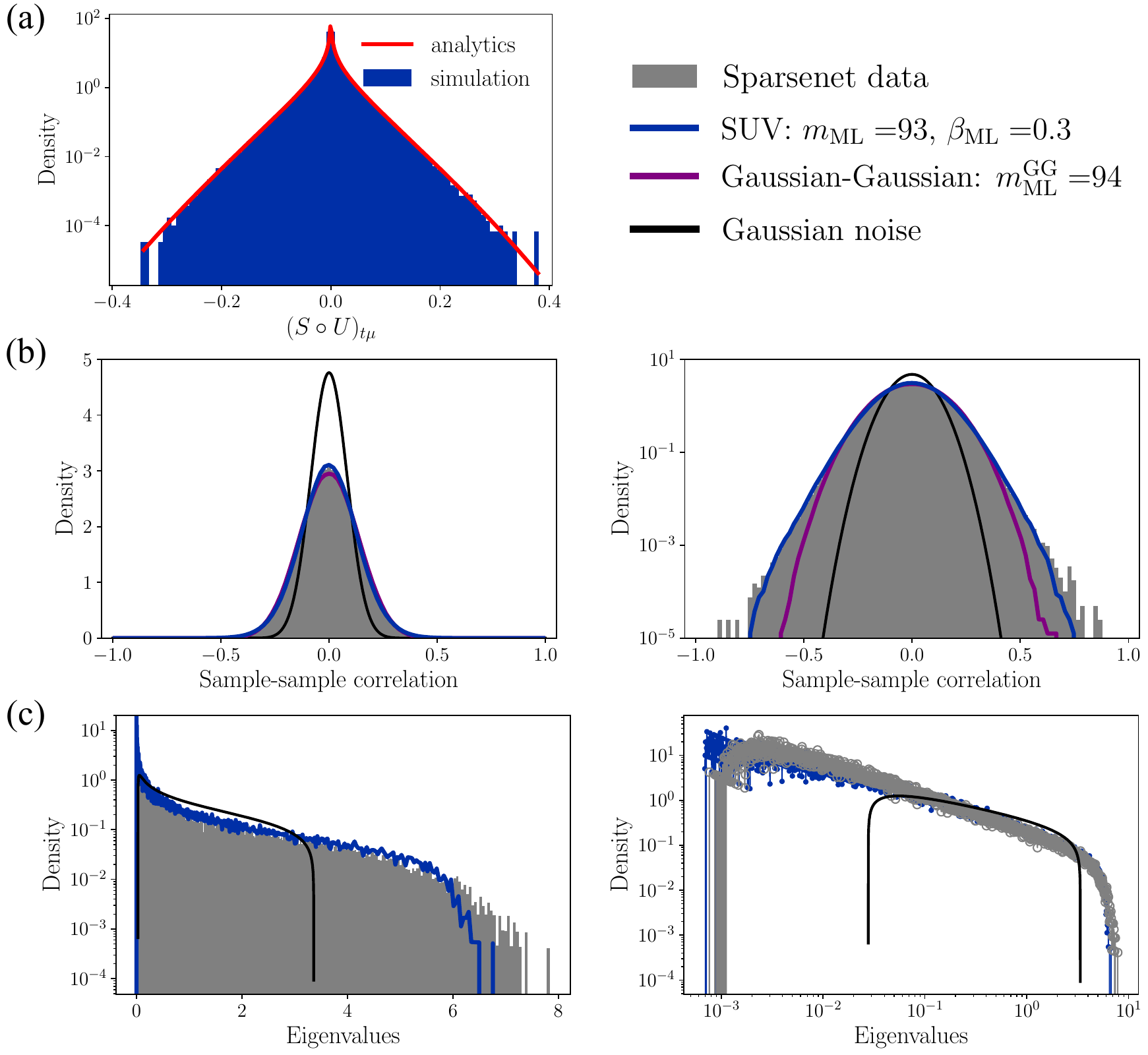}
\caption{{\bf Natural scene image sample statistics is well described by the SUV model.} We compare the sample-sample correlation and eigenvalue distributions of the sparsenet data set with predictions of the maximum-likelihood SUV model ($\ld_\mathrm{ML}= 93\pm1$, $\beta_\mathrm{ML}=0.30\pm0.01$). The {\em Sparsenet} data matrix consists of $T=100$ image patches each of size $N=12\times12=144$. To obtain the observable distribution (gray), we sample the data matrix multiple times, stated as number of trials below. The elements of the modulation matrix $\mathbf S$ are i.~i.~d. binary random variables taking values $\{\pm1\}$, each with probability $0.5$. (a) Distribution of elements of the mixing matrix $\mathbf S\circ\mathbf U$ of the maximum-likelihood model simulation (blue) and its analytic curve (red). (b) Comparison of sample-sample correlation of the sparsenet data (gray) and the SUV model (blue), on linear-linear (left) and log-linear scale (right). For reference we also show the distribution of correlations expected for pure noise data (black), and the Gaussian-Gaussian latent feature model (purple).  (c) Comparison of the eigenvalue density of the sparsenet sample-sample correlation matrix (gray) and the model prediction (blue), on log-linear scale (left) and log-log scale (right). For reference we show the Mar\v cenko-Pastur distribution (black line with sampling ratio $T/N$). All curves from model simulations and sparsenet data were computed with $400$ trials.}\label{fig:fig4}
\end{figure*}

We now turn to the analysis of the statistics of natural scene images. Significant success has been achieved in studying the correlation statistics between different pixels of natural images~\cite{ruderman1997origins,PhysRevLett.73.814}. Here, we study the correlations between samples (patches from natural scene images) and the eigenvalues of the sample-sample correlation matrix, and make the comparison with predictions of the SUV random latent feature model.

Our data comes the {\em Sparsenet} data set~\cite{olshausen_data} used by Olshausen et al.~to train a model of receptive fields in the mammalian visual system~\cite{olshausen1996emergence}. The data set consists of ten natural images (of North American landscape) in grayscale and each image has been whitened. Specifically, we construct a $T\times N$ data matrix $\mathbf X$, containing $T$ flattened, square image patches of $N=12\times12=144$ pixels. The image patches are randomly sampled from the ten images of the data set.

We standardise the rows of the data matrix and compute its sample-sample correlation matrix Eq.~(\ref{eq:corr_matrix}). The distribution of the correlations obtained from sampling the data matrix $400$ times, is shown in gray in Fig.~\ref{fig:fig4}(b). Sampling the data matrix multiple times is required in order to achieve good sampling of the distribution tails. 
To show that this distribution captures non-trivial structure of the data we show the distribution of correlations for i.~i.~d. Gaussian noise data (black curve). The correlations in noise data follow a Beta distribution~\cite{Hotelling_1953}, whose expression we provide in Eq.~(\ref{eq:Beta_distr}) of Appendix~\ref{sec:noise_observables}. For large enough $N$, the curve approaches a Gaussian normal distribution with variance $1/N$. The distribution of noise correlations substantially differs from the correlations in the sparsenet data. Next, we model the sparsenet correlations with the SUV model.

The first step is to choose a modulation matrix $\mathbf S$. For this we note that the correlation distribution,  Fig.~\ref{fig:fig4}(b), is symmetric around zero, taking on both positive and negative correlation values. This is different from our prior realizations of the SUV model, which skewed correlations to positive values. To allow for this symmetry, we need to allow for significant negative correlations, which could come from some samples including the same feature with positive and negative weights. To achieve this in our SUV model, we make the following  choice of the modulation matrix:
\begin{align}
    s_{t\mu}\sim P(k)=
\begin{dcases*}
p
   & $k=1$\,, \\[1ex]
1-p
   & $k=-1$\,,
\end{dcases*}
\end{align}
such that each entry of $\mathbf S$ is an i.~i.~d. binary random number, which, with a probability of $p$ is equal to $+1$, and with a probability of $1-p$ equal $-1$. This modulation flips the sign of each entry of the coefficient matrix $\mathbf U$ with probability $1-p$, allowing for negative mixing weights. Since the  {\em sparsenet} correlations are nearly perfectly symmetric, we set $p=1/2$. We show the distribution of mixing matrix entries $(\mathbf S\circ \mathbf U)_{t\mu}$ obtained in a simulation in Fig.~\ref{fig:fig4}(a) in blue, together with the analytic curve derived in Appendix~\ref{app:mixing_coeff} and given by Eq.~(\ref{eq:mixing_matrix_distribution}) in red.

The second step is to fit the model parameters $\beta_U$ and $\ld$ to the distribution of the correlations (gray in Fig.~\ref{fig:fig4}(b)) using maximum likelihood (ML) estimation. We find as ML values $\ld_\text{ML}=93\pm1$, and $\beta_\text{ML}=0.30\pm 0.01$. The ML prediction made by the SUV model is shown in blue in Fig.~\ref{fig:fig4}(b). We find a near-perfect match between the {\em Sparsenet} correlations and our model, which is even accurate in the tails of the distribution (Fig.~\ref{fig:fig4}(b) right). As an aside, we note that if we instead fit the model on single trials of the data matrix, we find $\ld_\text{ML}=92\pm5$, and $\beta_\text{ML}=0.33\pm 0.05$. The small difference in parameter estimates is due to a systematic underestimation of the distribution tails which are not well sampled in single trials. 

Now, to  make a {\em  prediction}, we use the SUV model with parameters inferred from the distribution of correlations and compare the eigenvalue distribution of its correlation matrix to the {\em Sparsenet} data. We show the comparison in Figs.~\ref{fig:fig4}(c) and (d). We find a good match between the data (gray) and the model prediction (blue). For reference, we also show the distribution of eigenvalues expected for pure noise data (black curve), given by the famous Mar\v cenko-Pastur distribution~\cite{Marchenko_Pastur1967}, which we write in Appendix~\ref{sec:noise_observables} in Eq.~(\ref{eq:MP_distr}).

Finally, we ask what the good match of the SUV model tells us about the underlying structure of the data. First, we note that the ML parameter values $\ld_\text{ML}\approx93$ and $\beta_\text{ML}\approx0.3$ imply that for each sample (image patch), only a relatively small number of the available total number of features is used for mixing, and the ML model is thus of sparse mixing type. Note that {\em sparse} in the name of the original data set refers to the structure of activity of the neural network used to analyse it. It is thus encouraging to see that the data themselves can be modeled as sparse.

Second, we point out that, with the inferred parameters, the overall dataset is not particularly low-dimensional (even if each patch is), since the number of features, $\sim 93$, is similar to the number of variables, 144. Thus we check whether the data could also be fit by a simpler, low-rank model. For this, we fit the correlation distribution of the Gaussian-Gaussian model on the correlation data and find an ML value of $\ld^\text{GG}_\text{ML}= 94\pm1$ closely matching the ML value determined for the SUV model.
In Fig.~\ref{fig:fig4} (b), we show the ML correlation curve of the Gaussian-Gaussian model in purple. The curve is a reasonably good match in the bulk of the distribution, but shows deviations in the tails, in contrast to the SUV model. We conclude that the most suitable model for the data is one which does not assume a pure low-rank structure in the classical sense, but one that introduces low dimensionality through sparse mixing of features. This is probably why data decomposition methods like Independent Component Analysis and sparse coding when applied to these natural images yield latent features which are interpretable as receptive fields~\cite{van1998independent,olshausen1996emergence}.

\section{Discussion}
We have presented a  random latent feature model for structured data, based on linear mixing of latent features,  Eq.~(\ref{eq:model}). We call this the SUV model. Key ingredient in our model is the statistical dependence of the mixing coefficients. In different parameter regimes, this dependence gives rise to different patterns, commonly found in natural data sets, from clusters, to sparse models. As an example, we showed that the SUV model is an {\em excellent} model for the sample-sample statistics of natural images, requiring fitting of just two hyperparameters: the size of the feature set and the average number of features used to generate any particular sample. We are not aware of similar simple, analytical generative models that can approximate such statistics in this or other natural datasets---usually fitting of many parameters, such as synaptic weights, is required for a similar accuracy~\cite{saremi_sejnowksi2012,schneidman2006weak,bialek2012statistical,gao2016linear,Pandarinath_etal_2018,sohl2015deep}.

A central point emerging from our analysis is that the qualitative structure  of the distribution of various observables can inform us about a class of models best suited for representing the data. For example, in the case of {\em Sparsenet} data, we observed that quantitatively small differences in the tails of the distribution of sample-sample correlations suggest that sparse mixing of features is more appropriate to describe this dataset  than more traditional low-rank models. For data analysis, this means that decomposition schemes like Independent Component Analysis~\cite{Bell1995AnIA,HYVARINEN2000411}, non-negative matrix factorization~\cite{lee1999learning}, and similar sparse coding models, are more natural to apply to the data---and are more likely to yield physically and biologically interpretable latent features---than other decomposition methods, such as Principal Component Analysis~\cite{shlens2014tutorial}. Similarly, a characteristic feature of our model is the aster structure of the data scatters, cf.~Fig.~\ref{fig:fig1}(e,f). These data are locally low dimensional, but globally they are not. Since such structures, indeed, occur  in data and models of neural activity~\cite{Berry2020ClusteringON,mlynarski2022efficient}, one should question whether models based on low-dimensional neural activity manifolds are the best model for such datasets. 

In this article, we have focused mostly on sample-sample correlations. This is because, in our opinion, these statistics do not receive enough attention in the literature. However, non-zero parts of spectra of sample-sample and variable-variable correlations are the same up to a normalization constant, and the SUV model can equally be applied to the variable-variable statistics. Thus our analysis is more general than it may seem.

The SUV model, Eq.~(\ref{eq:model}), is exceedingly simple, and it is unclear {\em a priori} why it should be so effective in modeling {\em Sparsenet} data.  We believe that there are two reasons for this. First, note that we assumed  the latent features to be Gaussian i.~i.~d. vectors,  which is certainly a bad model for receptive fields in \cite{olshausen1996emergence}. However, the feature structure becomes practically irrelevant to sample-sample correlations as long as the individual features are uncorrelated either due to the central  limit theorem at  large enough $N$, or due to features forming a sparse, overcomplete code, as in \cite{olshausen1996emergence}. Second, in our model, sparsity emerges as a consequence of the Dirichlet distribution in a particular parameter regime. This is analytically tractable, but is also unlikely to be a precise model of the world. Yet Olshausen et al.~\cite{olshausen1996emergence} showed that different ways of imposing sparsity on their model resulted in quantitative similar conclusions. Thus it is natural to expect that the details of the sparseness mechanism in the generative model would not be especially important either. The insensitivity of our model to the feature and the sparsity structures point to a notion of universality---a yet undiscovered class of models with quantitatively similar sample-sample correlations. Analysis of this class would need to be done using RMT approaches \cite{Potters_Bouchaud_2020}, and would parallel other  universality classes discovered within the theory. Note, however, that traditional universal results in RMT emerge from matrices with i.~i.~d.\ entries, and tools for analysis of statistically dependent mixing coefficients are not well developed. Therefore, we hope that the need to analyse the SUV or similar models will lead to  advances in the RMT methods as well.

\begin{acknowledgments}
IN was supported, in part, by the Simons Foundation Investigator award, NSF grant PHY/201052, and NIH grants 1R01NS099375 and 2R01NS084844.
\end{acknowledgments}



\providecommand{\noopsort}[1]{}\providecommand{\singleletter}[1]{#1}%
%

\appendix

\section{Computation of data mean and variance}\label{app:variance_GaussDirichlet}
\begin{figure}
\includegraphics[width=8.4cm]{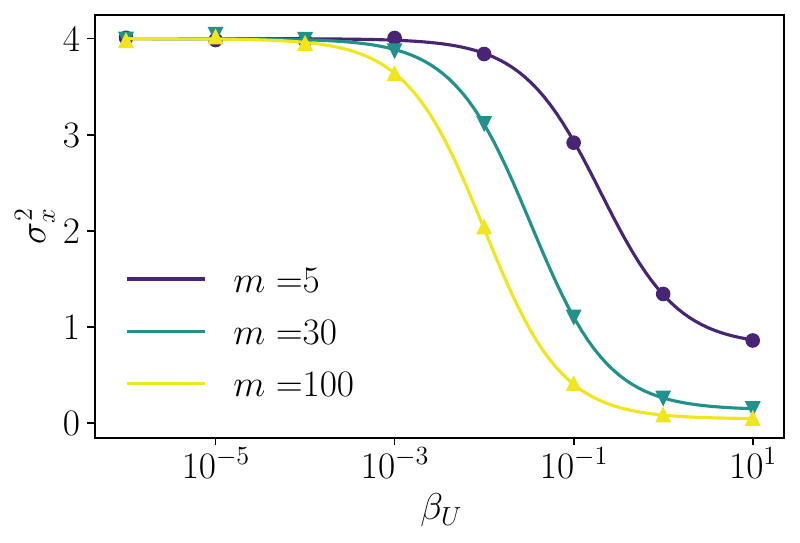}
\caption{{\bf Comparison of the analytic (solid lines) and simulated (markers) values of the data variance.} Dirichlet-Gaussian model with trivial modulation matrix, $\sigma_V^2=4$ for $\ld=5,30$, and $100$ over a range of $\beta_U$ values. Simulated values are averages over $3.6\times 10^4$ independent model realizations.}\label{fig:S1}
\end{figure}

We compute the theoretical mean and variance of data from the Dirichlet-Gaussian model with a trivial modulation matrix ($s_{t\mu}=1$), and i.~i.~d. Gaussian latent features. Starting from the model definition in Eq.~(\ref{eq:model}), we see that the mean, Eq.~(\ref{eq:mean_x}), and variance, Eq.~(\ref{eq:var_x}), are:
\begin{align}
    \mu_x&=\sum_\mu\mathbb E[u_\mu v_\mu]=\sum_\mu\mathbb E[u_\mu]\mathbb E[v_\mu]\,,\label{app:mean_x}\\
    \sigma^2_x&=\sum_\mu\text{var}(u_\mu v_\mu)+\sum_{\mu\neq\nu}\text{cov}\left(u_\mu v_\mu,u_\nu v_\nu\right)\,,\label{app:var_x}
\end{align}
where we imposed trivial modulation.
We note that the expression for the variance of the product of two independent random variables is
\begin{align}\label{app:var_prod}
    \text{var}\left(u_\mu v_\mu\right)&=(\sigma_u^2+\mu_u^2)(\sigma_v^2+\mu_v^2)-(\mu_u\mu_v)^2\,.
\end{align}
To compute the first term on the right-hand side of Eq.~(\ref{app:var_x}), we use Eqs.~(\ref{eq:Dirichlet_mean&variance}), (\ref{eq:DG_model}) and (\ref{app:var_prod}) to find
\begin{align}
    \text{var}\left(u_\mu v_\mu\right)&=(\sigma_u^2+\mu_u^2)(\sigma_V^2+\mu_V^2)-(\mu_u\mu_V)^2\nn\\
    &=(\sigma_u^2+\mu_u^2)\sigma_V^2=\left(\frac{1-\ld^{-1}}{\ld(1+\ld \beta_U)}+\frac{1}{\ld ^2}\right)\sigma_V^2\,.
\end{align}
The second term on the right-hand side of Eq.~(\ref{app:var_x}) vanishes for $\mu\neq\nu$, since
\begin{align}
    \text{cov}\left(u_\mu v_\mu,u_\nu v_\nu\right)&=\mathbb E\left[(u_\mu v_\mu-\mathbb E(u_\mu v_\mu))(u_\nu v_\nu-\mathbb E(u_\nu v_\nu))\right]\nn\\
    &=\mathbb E[u_\mu v_\mu u_\nu v_\nu]-\mathbb E[u_\mu u_\nu]\mathbb E[v_\mu v_\nu]=0\,,
\end{align}
where we have used statistical independence of $u$ and $v$, and the fact that $v$ has zero mean. Therefore, the expression for the variance is
\begin{align}
    \sigma^2_x=\ld\,\text var(u_\mu v_\mu)=\left(\frac{1-\ld^{-1}}{1+\ld\beta_U}+\frac{1}{\ld}\right)\sigma_V^2\,.
\end{align}
We verify this expression numerically in Fig.~\ref{fig:S1} (a).

\noindent

\begin{figure}
\includegraphics[width=8.4cm]{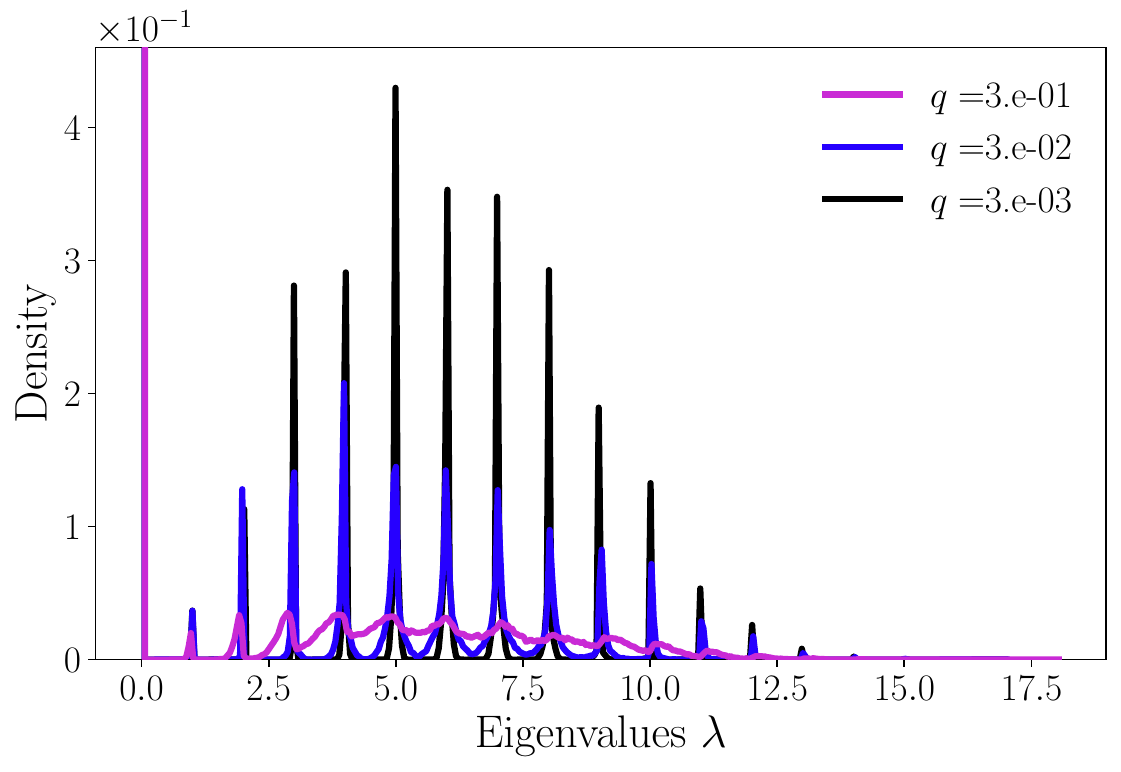}
\caption{{\bf Eigenvalue density spikes in the clusters regime of the Dirichlet-Gaussian model are resolved by small sampling ratios:} We show this for $T=30$, $\ld=5$, $\beta_U=4\times10^{-8}$, and $\sigma_V^2=1$. The resolution of the spikes increases as the sampling ratio $q=T/N$ decreases. Averages are computed from $2\times 10^4$ independent realizations.}\label{fig:S2}
\end{figure}

\begin{figure*}
\includegraphics[width=17.4cm]{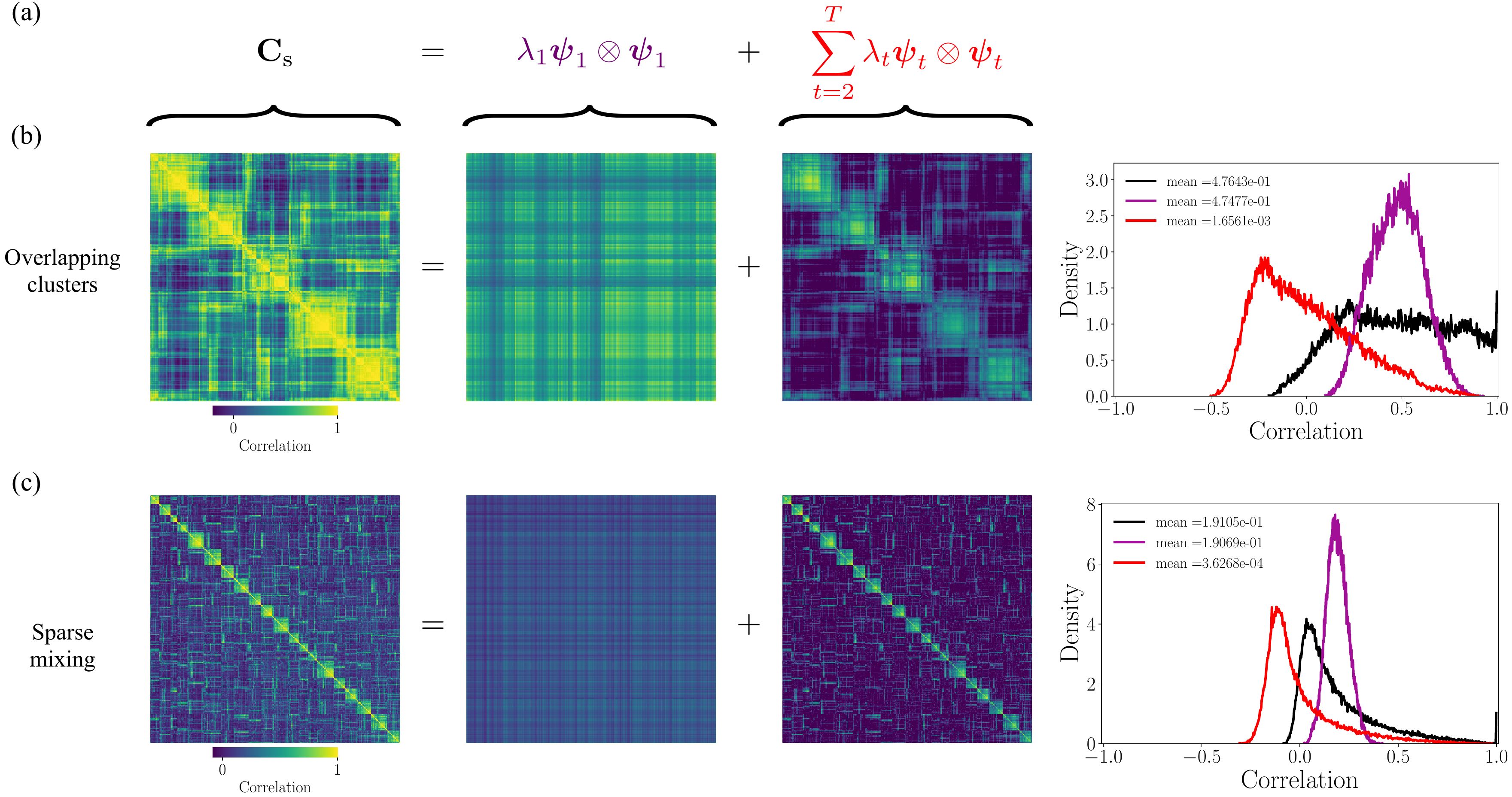}
\caption{{\bf Decomposition of the correlation matrix of the Dirichlet-Gaussian model.} We show this for a single realization with $T=360$. (a) General decomposition of the correlation matrix (black) into contributions of the top-ranked eigenmode $\boldsymbol{\psi}_1$ (purple) and the sum of the remaining eigenmodes $\boldsymbol{\psi}_{2,\ldots,T}$ (red). The contributions are given by the outer product $\otimes$ of eigenmodes, weighted by the eigenvalues. (b) The overlapping clusters model with $N=100$, $\ld=5$ and $\beta_U=0.4$. The distributions of matrix elements of the different contributions are shown on the right in their respective color. The numerical mean values are also stated. (c) The sparse mixing model with $N=1000$, $\ld=20$ and $\beta_U=0.15$. To the extent that the mean of the matrix elements of the first mode is essentially indistinguishable from the means of the overall ECM, and the means of all other eigenvalue contributions are nearly zero, this shows that the mean correlation in the ECM is captured by the top-ranked eigenmode.}\label{fig:S3}
\end{figure*}

\section{Analytic distribution for the mixing matrix with binary modulation}\label{app:mixing_coeff}
We derive an analytic expression for the distribution of mixing coefficients $(\mathbf S\circ \mathbf U)_{t\mu}$ for the case when the rows of the coefficient matrix $\mathbf U$ are given by Dirichlet random vectors as defined in Eq.~(\ref{eq:DG_model}), and each element of $\mathbf S$ is an i.~i.~d. binary random variable, equal to $+1$ with probability $p$, and $-1$ with probability $1-p$:
\begin{align}
    s_{t\mu}\sim P(k)=
\begin{dcases*}
p
   & $k=1$\,, \\[1ex]
1-p
   & $k=-1$\,.
\end{dcases*}
\end{align}
To obtain the distribution of individual entries of the mixing matrix, we marginalise the Dirichlet distribution Eq.~(\ref{eq:Dirichletdistr}) over all, but a single component. It is well-known that this results in the Beta distribution 
\begin{align}
    &\int du_1...du_{\ld-1}\delta\Big(1-\sum_\mu u_\mu\Big){\rm Dir}(\{u_\mu\}; \beta_U)\\
    &=\text{Beta}(u_\ld;\beta_U,(\ld-1)\beta_U)\,.
\end{align}
Since the hyperparameter $\beta_U$ of the Dirichlet distribution in Eq.~(\ref{eq:Dirichletdistr}) is the same for all variables $u_\mu$, the result of marginalising is independent of which of the $\ld-1$ directions we choose to marginalise over. All elements of the coefficient matrix thus follow the same Beta distribution:
\begin{align}
    u_{t\mu}\sim \text{Beta}(u_{t\mu};\beta_U,(\ld-1)\beta_U)\,.
\end{align}
To obtain the elements of the mixing matrix, the coefficients are modulated by $\mathbf S$, flipping their sign with a probability of $1-p$. To account for the modulation, we have to weight the distribution of positive and negative elements of the mixing matrix by $p$ and $1-p$, respectively. 
The distribution of mixing matrix elements
\begin{align}
        x \equiv (\mathbf S\circ \mathbf U)_{t\mu}
\end{align}
is, therefore, given by
\begin{align}\label{eq:mixing_matrix_distribution}
    \mathrm{pdf}(x) &=\Big[p H(x) + (1-p)H(-x)\Big] \mathrm{Beta}(|x|; \beta,(\ld-1)\beta)\,,
\end{align}
where $H$ is the Heaviside step function.

\begin{figure*}
\includegraphics[width=17.4cm]{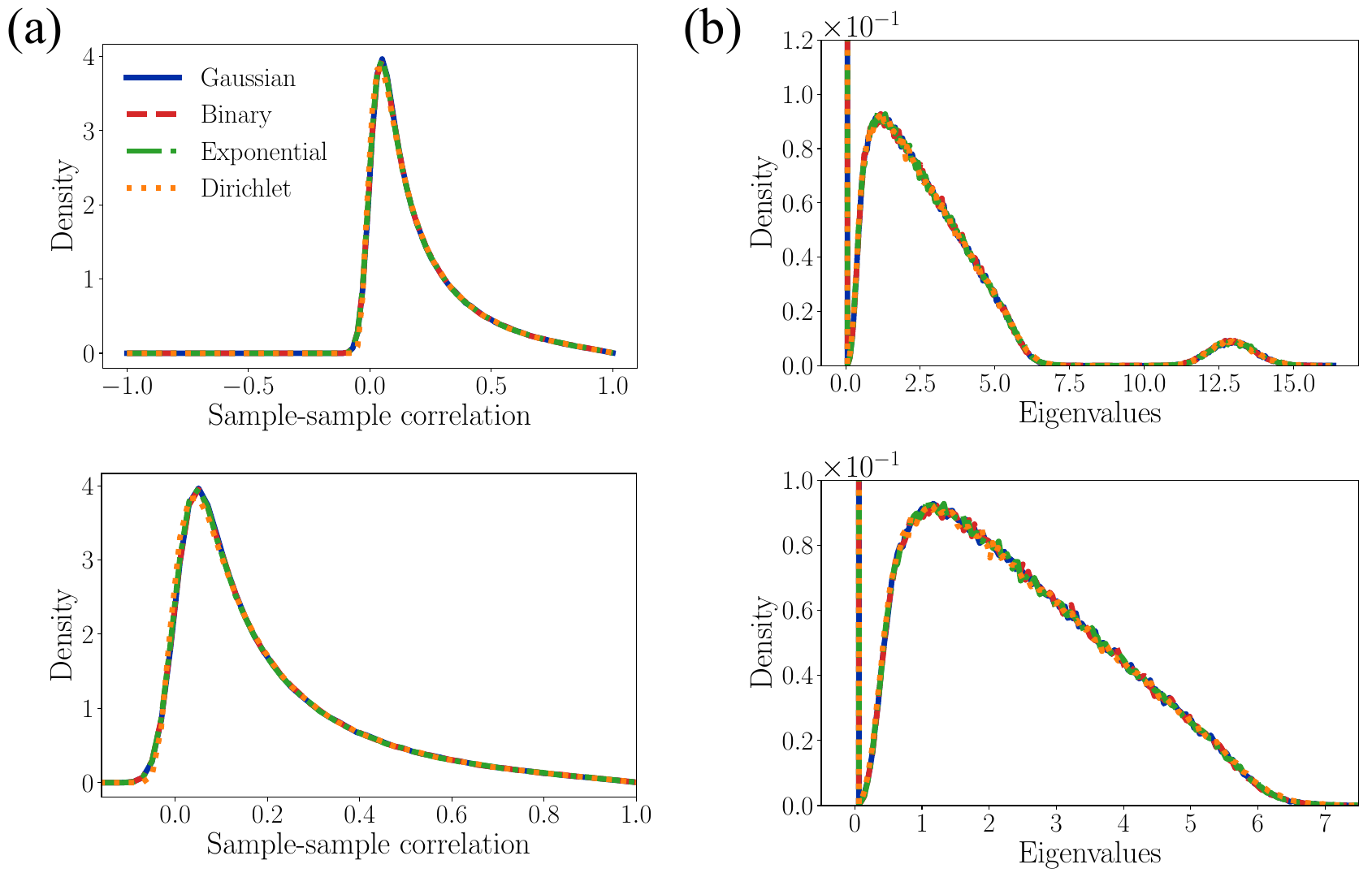}
\caption{{\bf Latent feature structure has almost no influence on the correlation and eigenvalue densities of the SUV model in the sparse mixing regime.} We use $T=60$, $N=1000$, $\ld=20$, $\beta_U=0.15$, and $s_{t\mu}=1$ for all entries. We test the following types of latent features $\mathbf V$: (blue line) i.~i.~d. Gaussian entries for reference, (dashed red) binary $\{\pm1\}$ i.~i.~d. Bernoulli entries with $p=0.5$ (dashed red), (dash-dotted green) i.~i.~d. exponentially distributed entries  with rate parameter $\lambda=1$, (dotted orange) Dirichlet random vectors with $\beta_V=5\times10^{-2}$. (a) Correlation densities computed from $800$ independent realisations of simulated data. Bottom plot is a zoom-in version of the top plot. (b) Eigenvalue densities computed from $2\times 10^4$ independent realisations of simulated data. Bottom plot is a zoom-in version of the top plot.}\label{fig:S4}
\end{figure*}

\section{Observable densities of i.~i.~d. Gaussian noise data}\label{sec:noise_observables}
In this section, we collect some known results on the distribution of correlations and eigenvalues of the correlation matrix for i.~i.~d. Gaussian noise data. Specifically, we consider a $T\times N$ matrix with i.~i.~d. Gaussian random variables as entries.

The correlation coefficient $r$ between samples is given by the product of two mutually independent random vectors of length $N$ with i.~i.~d. Gaussian entries and the distribution of the correlation coefficient was derived in~\cite{Hotelling_1953}. See also~\cite{Fleig_Nemenman_2022} for an extensive discussion. The coefficients are distributed according to a symmetric Beta distribution
\begin{align}\label{eq:Beta_distr}
    \mathrm{pdf}(r)&=\text{Beta}\left(r;\alpha,\alpha;\ell=-1,s=2\right)\\
    &=\frac{\Gamma\left(\alpha +\frac12\right)}{\Gamma\left(\frac12\right)\Gamma\left(\alpha\right)}(1-r^2)^{\alpha-1}\,,
\end{align}
where $\Gamma$ is the Gamma function, and $\ell$ and $s$ are the location and scale parameter, respectively. The parameters $\ell$ and $s$ are set such that the density is defined on the correlation interval $r\in [-1,1]$, and
\begin{align}
    \alpha=\frac{N-1}{2}\,.
\end{align}
The variance of the distribution with the scale $s=2$ is given by
\begin{align}
    \text{var}=\frac{s^2}{4(2\alpha+1)}=\frac{1}{2\alpha+1}=\frac1N\,.
\end{align}
For large values $N$, the distribution approaches the normal distribution with variance $1/N$.

Next, we turn to the eigenvalues of the sample-sample correlation matrix. The eigenvalues follow to the famous Mar\v cenko-Pastur distribution:
\begin{align}\label{eq:MP_distr}
    \rho(\lambda)=\frac{\sqrt{(\lambda_+-\lambda)(\lambda-\lambda_-)}}{2\pi q\lambda}
\end{align}
with
\begin{align}\label{eq:MP}
    \lambda_\pm = (1\pm\sqrt{q})^2,
\end{align}
and $q=T/N$. We note that, as we consider the eigenvalue spectrum of the sample-sample correlation matrix $\mathbf C_\mathrm{s}$, the sampling ratio $q=T/N$ is the inverse of the ratio $N/T$, which would be used for the variable-variable correlation matrix $\mathbf C_\mathrm{v}$, more commonly encountered in applications.

\section{Gaussian-Gaussian model}\label{app:Gaussian-Gaussian}
We give a brief summary of the Gaussian-Gaussian model, which has been analysed and discussed at length in~\cite{Fleig_Nemenman_2022}. This model can be used to describe pure low-rank structure in data. While in Ref.~\cite{Fleig_Nemenman_2022} the model included additive noise, we focus here only on the signal part of the model. The model's data matrix is given by the product of two probabilistic matrices:  
\begin{align}
    \mathbf X = \mathbf U \mathbf V\,,
\end{align}
with matrix entries given by i.~i.~d. zero mean Gaussian random numbers:
\begin{align}
   & u_{t\mu} \sim \mathcal N(0,\sigma_U^2)\,,\quad v_{\mu n} \sim \mathcal N(0,\sigma_V^2)\,,\label{eq:vars}\\
    &t=1,\ldots,T,\; \mu=1,\ldots,\ld,\; n=1,\dots,N.
\end{align}
In general, the random variables have variances $\sigma_U^2$ and $\sigma_{V}^2$, but throughout this article we set the variances to unity $\sigma_U^2=\sigma_V^2=1$.
The dimensions of the factor matrices $\mathbf{U}$ and $\mathbf{V}$ are $T\times \ld$ and $\ld\times N$, respectively, and the latent dimensionality is $\ld$.

\end{document}